\documentclass[aps,preprint,superscriptaddress,showpacs]{revtex4}%
\usepackage{amsfonts}
\usepackage{amsmath}
\usepackage{amssymb}
\usepackage{graphicx}
\usepackage{graphicx}%
\begin{document}
\title{Instantaneous processing of "slow light": amplitude-duration control, storage,
and splitting}
\author{R. N. Shakhmuratov}
\affiliation{Instituut voor Kern- en Stralingsfysica, Katholieke
Universiteit Leuven, Celestijnenlaan 200 D, B-3001 Leuven,
Belgium}\affiliation{Kazan Physical-Technical Institute, Russian
Academy of Sciences, 10/7 Sibirsky Trakt, Kazan 420029 Russia}
\author{A. A. Kalachev}
\affiliation{Kazan Physical-Technical Institute, Russian Academy
of Sciences, 10/7 Sibirsky Trakt, Kazan 420029 Russia}
\author{J. Odeurs}
\affiliation{Instituut voor Kern- en Stralingsfysica, Katholieke Universiteit Leuven,
Celestijnenlaan 200 D, B-3001 Leuven, Belgium}
\pacs{42.50.Gy}
\date{{ \today}}

\begin{abstract}
Nonadiabatic change of the control field or of the low-frequency coherence
allows for an almost instantaneous change of the signal field propagating in a
thick resonant absorber where electromagnetically induced transparency is
realized. This finding is applied for the storage and retrieval of the signal,
for the creation of a signal copy and separation of this copy from the
original pulse without its destruction.

\end{abstract}
\maketitle

A seminal idea of storage and retrieval of light pulses (SRLP) using a medium
with electromagnetically induced transparency (EIT) was proposed in Ref.
\cite{Lukin2000} by Fleischhauer and Lukin. Shortly after the proposal, SRLP
was experimentally demonstrated in ultracold Na \cite{Liu}, in hot Rb vapor
\cite{Phillips}, and in a solid (Pr:YSO) \cite{Turukhin}. Later, SRLP received
much attention in the framework of quantum computing and quantum memory (see,
for example, Ref. \cite{Kuzmich}). The core idea of such a storage is the
dark-state polariton \cite{Lukin2000}. This is a particular superposition of
the pulse amplitude and the atomic coherence created in a two-quantum process
by signal and control fields. The signal pulse drives the transition from the
populated ground state $\left\vert g\right\rangle $ to an excited state
$\left\vert e\right\rangle $, and the cw control field drives the transition
between state $\left\vert e\right\rangle $ and a metastable state $\left\vert
m\right\rangle $, which are initially not populated (see inset in Fig. 1).
Reducing adiabatically the amplitude of the control field to zero, one can
stop this polariton in a state that has zero amplitude for the signal pulse
component and a nonzero amplitude for the component containing the coherence
$g-m$. This coherence resembles a stand-still "spin wave" whose spatial shape
coincides with the spatial shape of the signal pulse if it would have zero
group velocity, $V=0$. An adiabatic increase of the control field amplitude
from zero back to its initial value retrieves the signal pulse from the "spin
wave", both propagating with group velocity $V\neq0$. Later, by numerical
simulations and simple analytical calculations it was shown \cite{Kochar01}
that the adiabatic change of the control field amplitude is not crucial for
SRLP. Even an abrupt change of the control field gives almost the same result:
the stand-still spin wave is formed after the control field is switched off
and then the signal pulse is retrieved after the control field is switched on again.

Recently we showed \cite{Shakh06} that actually the signal pulse entering an
EIT medium is transformed at the very beginning to the control field and
leaves the medium with group velocity $c$. Then an adiabaton \cite{Grobe} is
formed, which consists of a dip in the temporal profile of the control field
and a bump for the signal field, propagating together with reduced group
velocity $V$. For equal coupling constants and frequencies of the signal and
control fields, the sum of their energies is constant in any cross section of
the medium where the adiabaton is formed. Therefore, the bump and the dip in
the intensity profiles of the propagating fields exactly compensate each
other, and the energy of such an adiabaton is zero. This gives an important
clue for the understanding of the slow light propagation. First, the spatial
wave of the low frequency coherence (spin wave) is built up, propagating with
the slow group velocity $V$. Then this wave commands the control field to
produce the slow signal field, transforming a part of the control field energy
to and back from the signal field. Such a picture of the adiabaton suggests
that, if we would instantaneously switch off the control field, the slow
signal would not appear because no energy is available for the transformation
described above. The spin wave produced at the early stage of the adiabaton
formation is not destroyed but stops propagating because the source, i.e., the
control field, which drives the spin-wave propagation, is switched off.
Following this heuristic picture one can propose a new method to control a
signal field. Firstly, one can instantaneously change the amplitude or phase
of the control field. Since the control field is a source producing the slow
signal field, its change would result in an instantaneous change of the
amplitude or phase of the signal field. Secondly, one can almost
instantaneously change the spin wave by a short rf pulse. Because the spin
wave also controls the slow light production, its change would cause an
instantaneous change of the signal field as well. This gives new opportunities
to coherently control the properties of the radiation field.

In this paper, we develop a novel scheme of such a control of the signal field
and justify the heuristic arguments given above. In our scheme, first, we
double the intensity of the signal pulse by the instantaneous doubling of the
intensity of the control field. This operation also increases the group
velocity of the signal pulse by a factor of two, i.e., it becomes $2V$. Then
by a short rf pulse we transfer half of the population of the metastable state
$\left\vert m\right\rangle $ to another hyperfine level $\left\vert
M\right\rangle $ of the ground state atom, also metastable. This imprints a
snapshot of the spatial shape of the spin wave to state $\left\vert
M\right\rangle $. Since state $\left\vert M\right\rangle $ is supposed to be
not coupled to the other atomic states by the driving fields, this part of the
spin wave becomes stand-still. But what is left in state $\left\vert
m\right\rangle $ continues propagating with group velocity $2V$. Its
probability amplitude reduces by a factor of $1/\sqrt{2}$ and, therefore, the
intensity of the signal pulse drops by a factor of two back to its initial
value. After some delay time, which is long enough to ensure that the signal
pulse has already left the medium, we apply again an rf pulse to bring back
the atomic population from state $\left\vert M\right\rangle $ to state
$\left\vert m\right\rangle $, which has been already emptied by this time. The
control field acts on the appearing stand-still spin-wave in state $\left\vert
m\right\rangle $ such that the signal pulse is produced again and both, the
spin-wave and the signal pulse, travel with the group velocity $2V$.

The excitation scheme is shown in the inset in Fig. 1. Initially, two ground
state levels $m$ and $M$ are depopulated by optical pumping, and only state
$g$ is populated. The control field with the coupling amplitude (the Rabi
frequency) $\Omega$ is a cw propagating along coordinate $z$ in a thick
resonant sample. At time $t_{0}=0$, the signal pulse with coupling amplitude
$\Psi$ enters the sample and propagates in the same direction as the field
$\Omega$. It is assumed that $\Psi\ll\Omega$ and the spectral width of the
pulse $\Delta_{\Psi}$ is smaller than the width of the transparency window,
$\Delta_{T}=2\Omega^{2}/\gamma$, where $\gamma$ is the decay rate of the
coherence $g-e$, which is fast, $\gamma>2\sqrt{2}\Omega$. Since the signal
pulse is weak, we can apply the linear response approximation for the solution
of the Schr\"{o}dinger equation for the atomic state%
\begin{equation}
\left\vert \phi\right\rangle =C_{g}\left\vert g\right\rangle +C_{m}\left\vert
m\right\rangle +C_{M}\left\vert M\right\rangle +C_{e}\left\vert e\right\rangle
, \label{Eq24}%
\end{equation}
where $C_{M}=0$. State $\left\vert M\right\rangle $ need only be considered
when the rf pulse is applied. In this approach it is sufficient to consider
only the evolution of the amplitudes $C_{m}$ and $C_{e}$, which are described
by the equations%
\begin{equation}
\partial X(z,t)/\partial t=\Omega(z,t)Y(z,t), \label{Eq1}%
\end{equation}%
\begin{equation}
\partial Y(z,t)/\partial t=-\gamma Y(z,t)-\Omega(z,t)X(z,t)-\Psi(z,t),
\label{Eq2}%
\end{equation}
where $X=C_{m}$ and $Y=iC_{e}$. Here it is assumed that $C_{g}\approx1$ holds
with a small deviation of the order of $(\Psi/\Omega)^{2}\ll1$. If the
condition of the adiabatic following of the dark state, $\Delta_{\Psi}%
\ll\Delta_{T}$, is satisfied, an approximate solution of Eqs. (\ref{Eq1}%
),(\ref{Eq2}) can be easily found \cite{ShOd05}%
\begin{equation}
X(z,t)=-\Psi(z,t)/\Omega+..., \label{Eq3}%
\end{equation}%
\begin{equation}
Y(z,t)=-\Psi_{t}(z,t)/\Omega^{2}+..., \label{Eq4}%
\end{equation}
where $\Psi_{t}(z,t)=\partial\Psi(z,t)/\partial t$ and the dots stand for
terms that are at least $\Delta_{\Psi}/\Delta_{T}$ times smaller. The wave
equation for $\Psi$ is
\begin{equation}
\widehat{L}_{c}\Psi(z,t)=i\alpha C_{g}^{\ast}C_{e}\approx\alpha Y(z,t),
\label{Eq5}%
\end{equation}
where $\alpha$ is a coupling constant and $\widehat{L}_{c}$ is the
differential operator $\widehat{L}_{c}=\partial_{z}+c^{-1}\partial_{t}$, where
index $c$ stands for the group velocity of the wave. Substitution of the
solution (\ref{Eq4}) into Eq. (\ref{Eq5}) gives $\widehat{L}_{c}%
\Psi(z,t)=-(\alpha/\Omega^{2})\Psi_{t}(z,t)$, which can be transformed to
$\widehat{L}_{V_{1}}\Psi(z,t)=0$, where $V_{1}=(c^{-1}+\alpha/\Omega^{2}%
)^{-1}$ is the new group velocity. The solution of this equation is
$\Psi(z,t)=\Psi^{0}(t-z/V_{1})$, where $\Psi^{0}(t)$ is the amplitude of the
signal field at the input of the sample.

At time $t_{1}>0$, when the pulse is in the sample, we abruptly change the
amplitude of the control field: $\Omega(z,t)=\Omega\lbrack1+h\Theta
(t-t_{1}-z/c)]$, where $\Theta(t)$ is the Heaviside step function and we
choose $h=\sqrt{2}-1$ to double its intensity. This step-wise change of the
control field amplitude propagates in the sample with velocity $c$. Before it
arrives to the atoms with coordinate $z$, i.e., for $t<t_{z}=t_{1}+z/c$, their
amplitudes $X(z,t)$ and $Y(z,t)$ are described by Eqs. (\ref{Eq3}%
),(\ref{Eq4}). After $t_{z}$, the solution of Eqs. (\ref{Eq1}),(\ref{Eq2})
gives the following amplitudes
\begin{equation}
X(z,t)\approx-\frac{\Psi(z,t)}{\sqrt{2}\Omega}-\frac{K_{x}(t-t_{z})}{\sqrt
{2}\Omega}\Psi^{0}(t_{z}-z/V_{1}), \label{Eq7}%
\end{equation}%
\begin{equation}
Y(z,t)\approx-\frac{\Psi_{t}(z,t)}{2\Omega^{2}}+K_{y}(t-t_{z})\Psi^{0}%
(t_{z}-z/V_{1}), \label{Eq8}%
\end{equation}
where
\begin{equation}
K_{x}(\tau)=h\frac{\gamma_{+}e^{-\gamma_{-}\tau}-\gamma_{-}e^{-\gamma_{+}\tau
}}{\gamma_{+}-\gamma_{-}}\Theta(\tau), \label{Eq10}%
\end{equation}%
\begin{equation}
K_{y}(\tau)=h\frac{e^{-\gamma_{-}\tau}-e^{-\gamma_{+}\tau}}{\gamma_{+}%
-\gamma_{-}}\Theta(\tau), \label{Eq11}%
\end{equation}%
\begin{equation}
\gamma_{\pm}=\frac{\gamma}{2}\pm\sqrt{\frac{\gamma^{2}}{4}-2\Omega^{2}}.
\label{Eq12}%
\end{equation}
Here, the first and the second terms in the solution (\ref{Eq7}),(\ref{Eq8})
represent the main contributions originating from the nonhomogeneous term,
$-\Psi(z,t)$, in Eq. (\ref{Eq2}) and from the initial condition at $t_{z}$.
The omitted terms are at least $\Delta_{\Psi}/\Delta_{T}$ times smaller.

Before $t_{z}$, the atomic state $\left\vert \phi\right\rangle $ was close to
the dark state $\left\vert d\right\rangle =\cos\theta_{1}(z,t)\left\vert
g\right\rangle -\sin\theta_{1}(z,t)\left\vert m\right\rangle $, uncoupled from
the signal and control fields, where $\tan\theta_{1}(z,t)=\Psi^{0}/\Omega$
\cite{Arimondo96}. The abrupt change of the control field amplitude from
$\Omega$ to $\sqrt{2}\Omega$ makes this state coupled since it acquires a
particular component, which is the bright state $\left\vert b\right\rangle
=\sin\theta_{2}(z,t)\left\vert g\right\rangle +\cos\theta_{2}(z,t)\left\vert
m\right\rangle $, coupled to the signal and control fields \cite{Arimondo96},
where $\tan\theta_{2}(z,t)=\Psi(z,t)/\sqrt{2}\Omega$. The probability
amplitude of $\left\vert b\right\rangle $ is proportional to the signal field
amplitude $\Psi^{0}$. Therefore, the amplitudes of the transient part of the
solution, i.e., the second terms in Eqs. (\ref{Eq7}),(\ref{Eq8}), are
proportional to the signal field amplitude $\Psi^{0}$. Because of the fast
decay of the excited state, the atom terminates its evolution in a new dark
state with the mixing angle $\theta_{2}(z,t)$ [see Eq. (\ref{Eq7})].

With the solution (\ref{Eq8}), the propagation equation (\ref{Eq5}) can be
transformed to
\begin{equation}
\widehat{L}_{c}\Psi(z,t)=-A(z,t)\Psi_{t}(z,t)+\alpha K_{y}(t-t_{z})\Psi
^{0}(t_{z}-z/V_{1}), \label{Eq13}%
\end{equation}
where $A(z,t)=\beta_{1}+(\beta_{2}-\beta_{1})\Theta(t-t_{z})$, $\beta
_{1,2}=V_{1,2}^{-1}-c^{-1}$ and $V_{2}=(c^{-1}+\alpha/2\Omega^{2})^{-1}$ is
the new group velocity of the signal field after the jump of the amplitude
$\Omega(z,t)$. The solution of Eq. (\ref{Eq13}) for $t<t_{z}$ is
$\Psi(z,t)=\Psi^{0}(t-z/V_{1})$. For $t>t_{z}$ this solution changes to%
\begin{equation}
\Psi(z,t)=\Psi^{0}(T)+\Phi(z,t), \label{Eq14}%
\end{equation}%
\begin{equation}
\Phi(z,t)=\alpha%
%TCIMACRO{\dint \nolimits_{0}^{z}}%
%BeginExpansion
{\displaystyle\int\nolimits_{0}^{z}}
%EndExpansion
K_{y}(t-t_{z^{\prime}}-\frac{z-z^{\prime}}{V_{2}})\Psi^{0}(t_{z^{\prime}%
}-z^{\prime}/V_{1})dz^{\prime}, \label{Eq15}%
\end{equation}
where $T=\kappa(V_{2}/V_{1})[t-t_{c}-(z-z_{c})/V_{2}]$, $\kappa=(c-V_{1}%
)/(c-V_{2})$, and $t_{z^{\prime}}=t_{1}+z^{\prime}/c$. $z_{c}=V_{1}t_{c}$ is a
coordinate where at time $t_{c}=t_{1}c/(c-V_{1})$ the central part of the
signal pulse changes its velocity from $V_{1}$ to $V_{2}$. Taking the integral
(\ref{Eq15}) by parts and retaining only the two main terms, we obtain
\begin{equation}
\Phi(z,t)=\eta h\Psi^{0}(T)-\eta K_{x}(t-t_{z})\Psi^{0}(t_{z}-z/V_{1}),
\label{Eq16}%
\end{equation}
where $\eta=(c-V_{2})/c$. After a short time $t-t_{z}=\tau_{\gamma}%
\sim1/\gamma$, the function $K_{x}(t-t_{z})$ decays to zero. If $c\gg
V_{1},V_{2}$, then $V_{2}=2V_{1}$, $\eta\approx\kappa\approx1$, $z_{c}\approx
V_{1}t_{1}$ and hence, for $t>t_{z}+\tau_{\gamma}$ we have $\Psi
(z,t)=(1+h)\Psi^{0}(T)$ where $T=2\left[  t-t_{1}-(z-z_{c})/V_{2}\right]  $.
This means that after the abrupt change of the amplitude of the control field
the amplitude of the signal field also changes by the same factor $(1+h)$. In
such a way the ratio $\Psi(z,t)/\Omega(z,t)$ is conserved. Since
$C_{m}(z,t)\approx-\Psi(z,t)/\Omega(z,t)$, the spin wave also conserves its
amplitude and length. The latter coincides with the spatial length of the
signal pulse in the sample before the change of the control field. This length
is $l_{p}=V_{1}t_{p_{1}}$, where $t_{p_{1}}=1/\Delta_{\Psi}$ is the duration
of the signal pulse at the input. Meanwhile, the spin wave and the signal
field alter their group velocity from $V_{1}$ to $V_{2}$. Therefore, the
duration of the signal pulse shortens to $t_{p_{2}}=t_{p_{1}}V_{1}%
/V_{2}=t_{p_{1}}/2$ such that the spatial length, $l_{p}$, of the pulse and
the spin wave is conserved.

At time $t_{2}=t_{1}+\tau_{\gamma}+l_{p}/c$, all spatial components of the
spin wave and the signal pulse complete such a transformation. Following our
scheme of the signal field processing, at time $t_{2}$ we apply a short,
rectangular-shaped rf pulse, which drives resonantly the transition $m-M$ (see
inset in Fig. 1). The wavelength of the rf pulse is much greater than the
spatial length, $l_{p}$, of the signal pulse. Therefore, we disregard its
spatial dependence. The evolution of the probability amplitudes $C_{m}$ and
$C_{M}$ of the atomic state $\left\vert \phi\right\rangle $, Eq. (\ref{Eq24}),
is described by the equations
\begin{equation}
\partial C_{m}/\partial t=iPC_{M}+i\Omega C_{e}, \label{Eq17}%
\end{equation}%
\begin{equation}
\partial C_{M}/\partial t=iPC_{m}. \label{Eq18}%
\end{equation}
where $P$ is the amplitude of the coupling $m-M$ with the resonant rf pulse.
We take $P\gg\Omega$ and choose the duration of this pulse $\tau_{rf}$ such
that it forms a so called $\pi/2$--pulse: $P\tau_{rf}=\pi/4$ (see, for
example, Ref. \cite{Eberly} for the definition). Before the rf pulse, we have
$C_{M}=0$ and $C_{m}=-\Psi^{0}/\Omega$. Since $\tau_{rf}\Omega\ll1$, we can
disregard the interaction with the control field during the rf pulse. Then, at
the end of the pulse, $t_{3}=t_{2}+\tau_{rf}$, we have $C_{m}=-\Psi^{0}%
\cos(P\tau_{rf})/\Omega=-\Psi^{0}/\sqrt{2}\Omega$ and $C_{M}=-i\Psi^{0}%
\sin(P\tau_{rf})/\Omega=-i\Psi^{0}/\sqrt{2}\Omega$. To simplify our
consideration, we assume that the transition $M-e$ is not allowed or far from
resonance. Therefore the presence of the coherence $m-M$ does not influence
the signal and the control fields. Only the change of the probability
amplitude $C_{m}$ introduces transients. They are described by Eqs.
(\ref{Eq1}),(\ref{Eq2}), where $\Psi(z,t)$ and $\Omega(z,t)$ are replaced by
$\sqrt{2}\Psi^{0}$ and $\sqrt{2}\Omega$ (these amplitudes were present before
the rf at $t_{2}\approx t_{3}$). At the end of the rf pulse, $t_{3}$, the
initial condition for an atom with coordinate $z$ is $X(z,t_{3})=-\Psi
^{0}(T_{3})/\sqrt{2}\Omega$ and $Y(z,t_{3})=-\Psi_{t}^{0}(T)|_{t=t_{3}}%
/\sqrt{2}\Omega^{2}$, where $T_{3}=(V_{2}/V_{1})[t_{3}-t_{1}-(z-z_{c})/V_{2}%
]$. After these modifications the solution of Eqs. (\ref{Eq1}),(\ref{Eq2}) is%
\begin{equation}
Y(z,t)\approx-\frac{\Psi_{t}(z,t)}{2\Omega^{2}}-K_{y}(t-t_{3})\Psi^{0}(T_{3}),
\label{Eq19}%
\end{equation}
with the initial condition $\Psi_{t}(z,t_{3})=\sqrt{2}\Psi_{t}^{0}%
(T)|_{t=t_{3}}$.

The solution of the wave equation (\ref{Eq5}) for $t<t_{3}$ is $\Psi
(z,t)=\sqrt{2}\Psi^{0}(T)$. For $t>t_{3}$, the r.h.s. of this equation changes
to Eq. (\ref{Eq19}). Then its solution transforms to $\Psi(z,t)=\sqrt{2}%
\Psi^{0}(T)-\Phi_{1}(z,t)$, where $\Phi_{1}(z,t)$ coincides with the function
$\Phi(z,t)$ in Eq. (\ref{Eq15}), if $t_{z^{\prime}}$ and $V_{1}$ are replaced
by $t_{3}$ and $V_{2}$, respectively. Within the same approximation adopted
for $\Phi(z,t)$, we obtain $\Phi_{1}(z,t)=h\eta\Psi^{0}(T)-\eta K_{x}%
(t-t_{3})\Psi^{0}(T_{3})$. Thus, after a short time $\tau_{\gamma}$, i.e., for
$t>t_{3}+\tau_{\gamma}$, and if $\eta\approx1$, we have $\Psi(z,t)=\Psi
^{0}(T)$. This means that after the switch off of the rf pulse the signal
field changes its amplitude to its original value, which was at the input of
the sample.

We allow the signal field, having resumed its original amplitude, to leave the
sample. The "spin wave", $C_{m}(z,t)$, accompanying the signal field and
propagating with group velocity $V_{2}$, vanishes at the sample end, $z=l_{s}%
$. Meanwhile, the snapshot of the signal field at time $t_{2}$ was imprinted
by the rf pulse to the probability amplitude $C_{M}(z,t_{3})=-i\Psi^{0}%
(T_{3})/\sqrt{2}\Omega$ of state $\left\vert M\right\rangle $. This amplitude
can be considered as another "spin wave" with zero group velocity (stand-still
wave). Following the procedure described in the introduction, we apply a
second rf pulse at time $t_{4}$ when the signal field has already left the
sample. The second rf pulse lasts six times longer than the first rf pulse,
i.e., $6\tau_{rf}$, such that it forms a $3\pi$-pulse: $P6\tau_{rf}=3\pi/2$.
The initial condition for this pulse is $C_{m}(z,t_{4})=0$ and $C_{M}%
(z,t_{4})=-i\Psi^{0}(T_{3})/\sqrt{2}\Omega$. According to Eqs. (\ref{Eq17}%
),(\ref{Eq18}), at the end of this rf pulse, $t_{5}=t_{4}+6\tau_{rf}$, we have
$C_{m}=\Psi^{0}(T_{3})\sin(P6\tau_{rf})/\sqrt{2}\Omega=-\Psi^{0}(T_{3}%
)/\sqrt{2}\Omega$ and $C_{M}=-i\Psi^{0}(T_{3})\cos(P6\tau_{rf})/\sqrt{2}%
\Omega=0$. The extra $2\pi$-rotation of the pseudospin $1/2$, corresponding to
the transition $m-M$, is necessary to obtain the proper sign for the final
value of $C_{m}$. Otherwise, if an rf pulse with $\pi$-area would be applied,
the $C_{m}C_{g}^{\ast}$ coherence would generate a signal field with a phase
that is opposite to the initial one. In our case this coherence generates a
field $\Psi$ with the same phase as before. To show this we solve Eqs.
(\ref{Eq1}),(\ref{Eq2}) with an arbitrary function $\Psi(z,t)$ and
$\Omega(z,t)=\sqrt{2}\Omega$ for the initial condition $Y(z,t_{5})=0$,
$X(z,t_{5})=-\Psi^{0}(T_{3})/\sqrt{2}\Omega$, and $\Psi(z,t_{5})=0$. The
solution for $Y(z,t)$ is%
\begin{equation}
Y(z,t)\approx-\frac{\Psi_{t}(z,t)}{2\Omega^{2}}+\frac{K_{y}(t-t_{5})}{h}%
\Psi^{0}(T_{3}). \label{Eq21}%
\end{equation}
Substituting $Y(z,t)$ to the wave equation (\ref{Eq5}), we obtain the solution%
\begin{equation}
\Psi(z,t)=\frac{\alpha}{h}%
%TCIMACRO{\dint \nolimits_{0}^{z}}%
%BeginExpansion
{\displaystyle\int\nolimits_{0}^{z}}
%EndExpansion
K_{y}(t-t_{5}-\frac{z-z^{\prime}}{V_{2}})\Psi^{0}(T_{3}^{\prime})dz^{\prime},
\label{Eq23}%
\end{equation}
where $T_{3}^{\prime}=(V_{2}/V_{1})[t_{3}-t_{1}-(z^{\prime}-z_{c})/V_{2}]$.
Approximately this integral is $\Psi(z,t)=\eta\lbrack\Psi^{0}\left(
T_{3-5}\right)  \Theta(t-t_{5})-\Psi^{0}\left(  T_{3}\right)  K_{x}%
(t-t_{5})/h]$, where $T_{3-5}=(V_{2}/V_{1})[t-t_{1}+t_{3}-t_{5}-(z-z_{c}%
)/V_{2}]$. If $\eta\approx1$, then for $t>t_{5}+\tau_{\gamma}$ we have
$\Psi(z,t)=\Psi^{0}\left(  T_{3-5}\right)  $ and the signal field is retrieved
from the spin coherence. Now we have a copy of the signal field in the sample
and the signal field outside the sample, both with the same amplitude and
duration. Fig. 1 shows a 3-d plot of the signal field evolution controlled by
the amplitude change of the control field and rf pulses. By numerical
simulations we verified our approximate solution and obtained a fair
agreement. It becomes almost perfect if our idealized solution is convoluted
with a Gaussian function, described in \cite{ShOd05}, which takes into account
a pulse broadening due to the narrowing of the EIT window with distance.

Summarizing, we found that the instantaneous change of the amplitude of the
control field produces an almost instantaneous change of the amplitude of the
signal field and it does not affect the amplitude of the low-frequency
coherence (spin wave). This change also results in the variation of the group
velocity and duration of the signal pulse. However, their product, which is
the spatial length of the pulse and the spin wave, does not change if both are
in the EIT sample when the change happens. The instantaneous change of the
spin-wave amplitude by a short rf pulse produces an instantaneous change of
the amplitude of the signal field without changing its group velocity and
duration. By a train of rf pulses the signal field can be split in two parts,
one of which can be temporarily stored in the sample. These findings can be
applied for information processing and storage, creation of a new type of
entangled states, if a signal field contains only one photon. For example, a
single photon can be split in two spatially and temporarily separated parts
for the preparation of time-bin qubits, which are of importance for quantum
communication \cite{Gisin}.

This work was supported by the FWO Vlaanderen and the IAP program of the
Belgian government.

\newpage\begin{figure}[ptb]
\resizebox{0.5\textwidth}{!}{\includegraphics{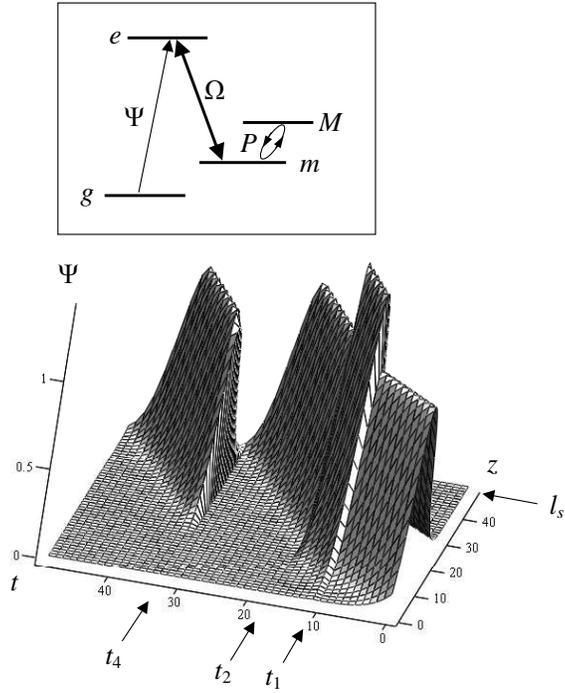}}\caption{3-d plot
showing the spatial-temporal evolution of the signal field. Coordinates $z$,
$t$ and amplitude $\Psi$ of the signal field are in arbitrary units. The
instances of the control, $t_{1}$, $t_{2}$, and $t_{4}$, are indicated by
arrows, as well as the end of the sample $z=l_{s}$. The excitation scheme of
the four-level atom is shown in the inset (see the text for details).}%
\label{fig:1}%
\end{figure}

\end{document}